\documentclass[preprints,article,accept,moreauthors,pdftex]{mdpi} 

\firstpage{1} 
\pubvolume{xx}
\issuenum{1}
\articlenumber{5}
\pubyear{2020}
\copyrightyear{2020}
\history{}

\usepackage{physics}
\usepackage[export]{adjustbox}
\usepackage{subcaption}
\usepackage{comment}
\usepackage{amsmath}
\usepackage{amssymb}

\newcommand{\Gaus}{\mathcal{G}}
\newcommand{\NC}{\mathcal{N}}
\newcommand{\kk}{k_0}

\DeclareMathOperator{\Var}{Var}

\Title{Scattering as a quantum metrology problem: a quantum walk approach}

\Author{Francesco Zatelli $^{1}$, Claudia Benedetti $^{1}$, and Matteo G. A. Paris $^{1,2}$,}

\AuthorNames{Francesco Zatelli, Claudia Benedetti and Matteo G. A. Paris}

\address{%
$^{1}$ \quad Dipartimento di Fisica 'Aldo Pontremoli', Universit\`a degli Studi di Milano, I-20133 Milano, Italy\\
$^{2}$ \quad INFN, Sezione di Milano, I-20133 Milano, Italy}
\corres{Correspondence: francesco.zatelli@gmail.com (FZ); claudia.benedetti@unimi.it (CB); matteo.paris@fisica.unimi.it (MGAP)}

\abstract{We address the scattering of a quantum particle by a one-dimensional barrier potential over a  set of discrete positions. We formalize the problem as a continuous-time quantum walk on a  lattice with an impurity, and use the quantum Fisher information as a mean to quantify the maximal possible accuracy in the estimation of the  height of the barrier. We introduce suitable initial states of the walker and derive the reflection and transmission probabilities of the scattered state.  We show that while the quantum Fisher information is affected by the width and central momentum of the initial wave packet, this dependency is weaker for the quantum signal-to-noise ratio.
We also show that a dichotomic position measurement provides a nearly optimal detection scheme. }

\keyword{Quantum walks; scattering; quantum metrology; quantum Fisher information;  optimal measurement.}

\begin{document}
\section{Introduction}
Since the Rutherford experiment \cite{ruthe}, scattering has played a central role in the study of unknown interactions in many fields of physics \cite{frankl, antipro, higgs}. At its core, a scattering experiment may be viewed as a parameter-estimation problem. 
Indeed, the  scattering potential can be modeled with a set of unknown parameters that characterize the evolution of the  quantum particles that impinge on it. 
 Estimating the value of those parameters then involves  measurements that are  performed on the scattered state, followed by a collection of outputs that are used to build  estimators for the parameters. If we consider scattering as an estimation problem, we can study the  maximum amount of information that can be extracted from a single measurement on the quantum system, and we can assess the performance of feasible detection schemes.  All these questions find answers in the theory of  local quantum estimation, which has the aim of quantifying the best precision of an estimation procedure \cite{helstr}. Indeed, in the past years, local quantum estimation theory has been applied to a variety of problems, such as estimation of the relevant parameters of a quantum structured baths \cite{gebbia20,hp20,Salari19,mirkin,wu}, graph and lattice properties \cite{seveso, fey,razzoli}, classical processes \cite{ben14}.
\par
  In this work we  analyze the one-dimensional scattering of a quantum particle from a potential barrier with the aim of inferring its height. The particle moves on a set of discrete positions, and it is thus described as a continuous-time quantum walk (CT QW) on the line with a central barrier. The barrier is implemented by a  detuning of the energy of the central site with respect to the other sites. 
As a matter of fact, the analysis of the evolution of a quantum walk  in the presence of a barrier is strongly connected with the study of 
 defects and impurities in  implementations of QW \cite{schreiber11, izaac13, li13, li15}. A detuning in the on-site energy of a site can be interpreted as  a defect, that influences the dynamics and the scattering properties of the walker.
 Understanding the role of imperfections is of fundamental importance for a realistic description of the QWs. In fact, knowing how   a  protocol or an algorithm \cite{defalco13,omar17,cattaneo18,kendon19,epl19} is affected by impurities and noise allows us to hinder or even  neutralize detrimental  effects.
\par
 Inspired by previous works on the discretization of continuous-systems \cite{tim04,tarasov}, we first derive scattered states on the infinite line of discrete positions.
  In order to consider physically relevant states for the walker, we initialize the particle in a Gaussian wave packet with central initial momentum $k_0$ and standard deviation $\sigma$. We  evaluate the transmission probability through the barrier and the maximum extractable information as a function of these two free parameters.  We 
  show that the quantum Fisher information (QFI) is strongly affected by the value of the initial central momentum of the walker, but only slightly by the initial spread of the wave packet. The quantum signal-to-noise ratio has a maximum corresponding to the  optimal value of the barrier height that can be better estimated.
    Finally, we consider a feasible measurement, i.e. a dichotomic position measurement, and we compare its Fisher information (FI) with the QFI. We show that this measurement is nearly optimal, i.e. its FI is close to the QFI in almost all the parameter space we consider.
\par
  The paper is organized as follows: in Section \ref{introqw} we introduce the concept of CTQW with inhomogeneous on-site energies and in Section \ref{introlqe} we briefly review the main concepts of local quantum estimation theory; In Section \ref{scatt} we introduce the free-particle scattering states and then we use them to build the physically relevant wave packets, whose transmission and reflection probabilities are derived. In Section \ref{est}   we compute the QFI for initial Gaussian wave packets and we compare  its value with the FI of a dichotomic position measurement.
  Finally, in Section \ref{concl} we draw our conclusions.
\section{Quantum walks with inhomogeneous on-site energies} \label{introqw}
%
%
%
%
A CTQW model  describes the evolution of a quantum particle over a discrete set of positions, continuously in time \cite{farhi,mulken}. 
It evolves in an $N$-dimensional Hilbert space with orthonormal basis states $\{\ket{j}\}_{j\in \mathbb{Z}}$ which represent the  positions that can be occupied by the walker. 
The Hamiltonian of a CTQW on the line with inhomogeneous on-site energies $\epsilon_j$ and  uniform couplings $J_0$ has the expression ($\hbar=1$):
\begin{equation}
    H=\sum_{j}\epsilon_j \ketbra{j}{j}-J_0\sum_{j}\Big(\ketbra{j}{j+1}+\ketbra{j+1}{j}\Big).
    \label{eqn:hamiltonian}
\end{equation}
Without loss of generality, we fix  $J_0=1$ , thus expressing time and the $
\epsilon_j$ in unit of $J_0$. 
If we set  $\epsilon_j=2\,\forall j$ we recover the graph Laplacian $L$, i.e. $H=- L$. It is worth mentioning that for the one-dimensional lattice,  $L$ represents the discretized version of  Laplace operator $\nabla^2$ and $-L$ is kinetic energy operator of a particle with mass $m=\frac{1}{2}$   constrained to a discrete set of positions \cite{wong16}. 
\par
Given a set of on-site energies $\{\epsilon_j\}$, it is  possible to separate the Hamiltonian into a kinetic   and a
 potential   operator, $L$ and $V$ respectively.  The Hamiltonian can thus be written as $ H=-L+V$ with:
  \begin{align}
 L=\sum_j\Big[-2\ketbra{j}{j}+\ketbra{j}{j+1}+\ketbra{j+1}{j}\Big]\quad \text{and} \quad V=\sum_{j}V_j\ketbra{j}{j}=\sum_{j}(\epsilon_j-2)\ketbra{j}{j}
 \label{decomp}
 \end{align}
 highlighting the fact that for $\epsilon_j=2\,\forall j$ the unperturbed Laplacian Hamiltonian is obtained. 
Due to the tridiagonal form of the matrix $H$, the eigenvalue equation $H|\psi^{(k)}\rangle=E_k|\psi^{(k)}\rangle$ can be 
recast in the form of a three-term recurrence relation. 
%
By explicitly writing  $H$ in terms of Laplacian and potential parts, and projecting into a basis state $\ket{j}$, we
obtain $ \bra{j}-L + V |\psi^{(k)}\rangle=E_k \bra{j}\psi^{(k)}\rangle$ 
and the recurrence relation:
\begin{align}
 -\psi_{j+1}^{(k)}+2\psi_j^{(k)}-\psi_{j-1}^{(k)}+V_j \psi_j^{(k)}=E_k \psi_j^{(k)},
         \label{eqn:eigenseq}
\end{align}
where $|\psi^{(k)}\rangle= \sum_{j}\psi_j^{(k)}\ket{j}$.
Eq. \eqref{eqn:eigenseq} is easily identifiable with  the discretization in  position basis of the time-independent Schrödinger equation for a particle of mass $m=\frac{1}{2}$.
\par
In analogy with the continuous case, we   introduce the momentum states as the Fourier series of the countable orthonormal set of position eigenstates.
In particular, we define the momentum  state $\ket{k}$
 through discrete-time Fourier transform (DTFT):
\begin{align}
    \ket{k}&=\frac{1}{\sqrt{2 \pi}}\sum_{j \in \mathbb{Z}} e^{ik j}\ket{j}, \quad k \in (-\pi, \pi] \label{kmoment}\\
    \ket{j}&=\frac{1}{\sqrt{2\pi}}\int_{-\pi}^{\pi}e^{-ikj}\ket{k}\dd{k}, \quad j \in \mathbb{Z}.
    \label{eqn:qft_infinite}
\end{align}
If no external potential is considered, i.e. $V_j=0\,\forall j$,  the states $\{\ket{k}\}$ are solutions to equation \eqref{eqn:eigenseq} with  $\psi_j^{(k)} = e^{ikj}$, and corresponding energies 
$ E_k=2-2\cos(k)$.
The  dispersion relation implies that the phase velocity $v_p$ and the group velocity $v_g$ are:
\begin{equation}
    v_p=\frac{E_k}{k}=\frac{2-2\cos(k)}{k}, \quad v_g=\pdv{E_k}{k}=2\sin(k).
    \label{eqn:velocities}
\end{equation}
Thus, the momentum states \eqref{kmoment} are the discretization of the plane waves with the dispersion relation typical of the tight-binding models \cite{simon}. We  identify these states as {\it free particle states} because, 
in analogy with the continuous case,  plane waves are the eigenstates of a purely kinetic Hamiltonian. This  suggests that the separation of the QW Hamiltonian into a kinetic term and a potential one is indeed meaningful. In the following we are going to introduce an obstacle, i.e. an external potential that causes an inhomogeneity on the on-site energies.
%
\section{Tools of local quantum estimation theory}\label{introlqe}
Before analyzing the  QW scattering from a barrier, we review few key concepts 
in the theory of local quantum estimation. Consider a sample of $M$ independent 
outcomes of a measurement $\{x_1, x_2, \dots, x_M\}$ drawn from the probability distribution $p(x|\Delta)$, where $\Delta$ is an unknown parameter we wish to estimate. 
The Cram\`er-Rao (CR) inequality imposes a lower bound on the variance of any unbiased estimator $\hat{\Delta}(\{x_1, x_2, \dots, x_M\})$ for such parameter:
\begin{equation}
    \Var(\hat{\Delta})\geq\frac{1}{M F(\Delta)}
    \label{eqn:cramerrao}
\end{equation}
where $F(\Delta)$ is the {Fisher information}, defined as:
\begin{equation}
    F(\Delta)=\int\qty(\pdv{\ln p(x|\Delta)}{\Delta})^2 p(x | \Delta) \dd{x}=\int\qty(\pdv{p(x|\Delta)}{\Delta})^2 \frac{1}{p(x | \Delta)} \dd{x}.
    \label{eqn:fi}
\end{equation}
The quantum version of the CR bound is derived by generalizing the concept of FI. 
This is done by maximizing the FI over all possible measurements, and the obtained quantity is called quantum Fisher information $H(\Delta)$. A detailed derivation of the QFI can be found in \cite{parislqe}.
The quantum CR bound takes the  following form:
\begin{equation}
    \Var(\hat{\Delta})\geq\frac{1}{M H(\Delta)}.
    \label{eqn:quantumcramerrao}
\end{equation}
and follows from the inequality $F(\Delta)\leq H(\Delta)$ , which provides the basis for the identification of the QFI with the ultimate bound the precision of an unbiased estimator. The aim of local quantum estimation theory is to determine the maximum extractable information from a quantum probe, whose state  depends on the value of the parameter. 
If only pure states are considered as probes,  
i.e. a parameter-dependent family of quantum  states $\ket{\psi_{\Delta}}$, the QFI can be explicitly written as \cite{parislqe}:
\begin{equation}
    H(\Delta)=4\qty[\braket{\partial_\Delta\psi_\Delta}{\partial_\Delta\psi_\Delta}-\abs{\braket{\psi_\Delta}{\partial_\Delta \psi_\Delta}}^2],
    \label{eqn:qfipure}
\end{equation}
where $\ket{\partial_\Delta\psi_\Delta}$ represents the derivative of the state with respect to the parameter $\Delta$.
A suitable figure of merit that can be used in order to evaluate the estimability of a parameter is the quantum signal-to-noise ratio (QSNR)
\begin{equation}
    R(\Delta)=\Delta^2 H(\Delta)\,,
    \label{qsnr}
\end{equation}
which provides an upper bound to the signal-to-noise ratio $\hat{\Delta}^2/  \Var(\hat{\Delta})$ of any detection scheme.
%
%
\section{Scattering in the presence of  an obstacle}\label{scatt}
Let us now  consider a situation where there is  an obstacle placed in the middle of the  chain. The obstacle, or barrier, has the width of a single site, i.e.
all sites have the same energy $\epsilon_j=2$, except for  the central one $\ket{0}$
 which has a detuning $\Delta$, such that $\epsilon_0=2+\Delta$.
Thus, the Hamiltonian defined in Eq. \eqref{eqn:hamiltonian} is modified by placing the obstacle at $j = 0$
and  it  becomes:
\begin{equation}
    H=\sum_{j\in \mathbb{Z}}\Big(2\ketbra{j}{j}-\ketbra{j+1}{j}-\ketbra{j}{j+1}\Big)+\Delta \ketbra{0}{0}.
    \label{eqn:hamiltonian_scattering}
\end{equation}
The site $j=0$ has on-site energy $\epsilon_0=2+\Delta$ or, alternatively said, potential $V_0=\Delta$. 
In order to study the scattering properties of such model, we start by deriving the scattering states.
\subsection{Scattering states}
\begin{figure}[t]
    \centering
        {\includegraphics[width=0.95\textwidth]{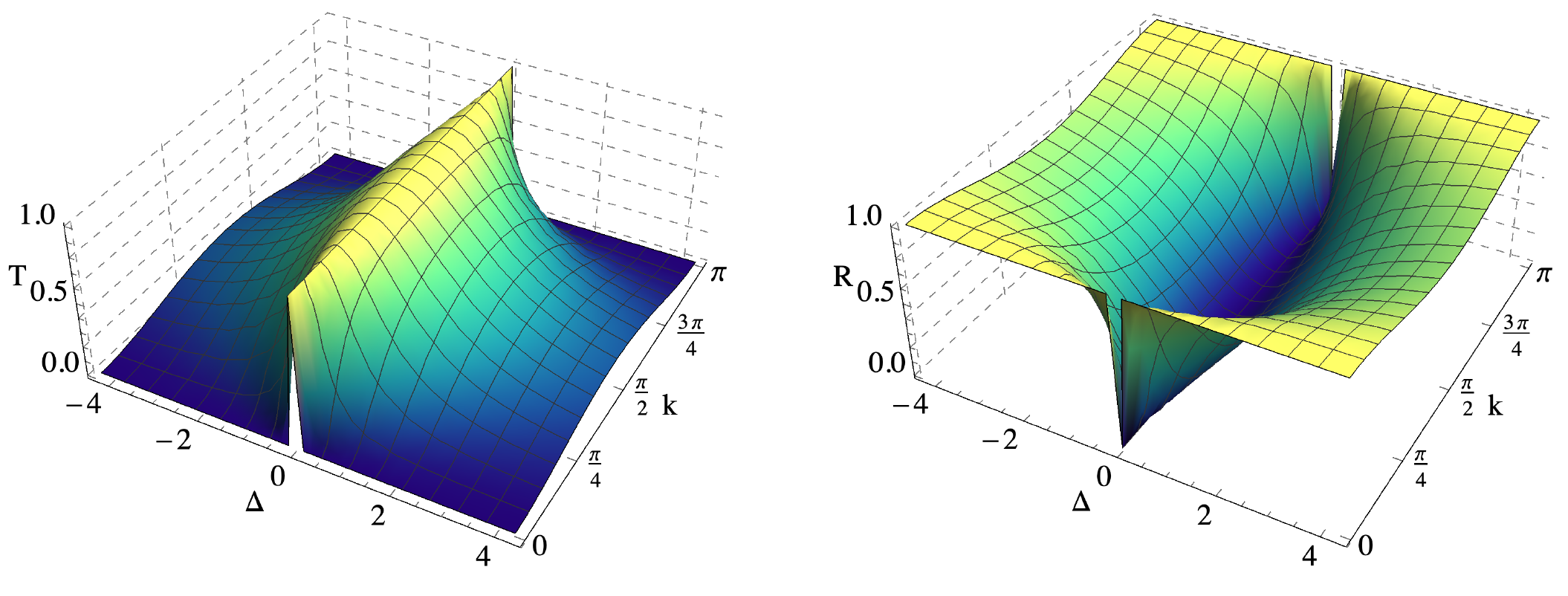}}
    \caption{Transmission and reflection coefficient $T$ and $R$ a as a function of $\Delta$ and $k$.  } 
      \label{fig:trcoefficient}
\end{figure}

Scattering states for one-dimensional systems in the continuous-space case 
are know for a variety of potentials \cite{Griffiths}. We now want to derive such states
 for the discrete system under consideration.
The generic stationary scattering state $\ket{\psi_s}$ with fixed momentum $k$ can be written as a linear combination of free particle states, namely:
\begin{equation}
    \braket{j}{\psi_s}=\begin{cases}
        Ae^{ikj}+Be^{-ikj}, & j\leq 0\\
        Ce^{ikj}, & j\geq 0
    \end{cases}\,,
    \label{scatstates}
\end{equation}
where the terms proportional to $A$, $B$ and $C$ correspond to the the incident, the reflected and the transmitted wave respectively.
The coefficients are calculated imposing that the two parts of the state (before and after the obstacle) are properly connected at $j=0$, i.e. by discretizing the continuity conditions, and using the recurrence relations \eqref{eqn:eigenseq}, i.e. $\langle{-1}|{\psi_s}\rangle-\Delta\langle{0}|{\psi_s}\rangle+\langle{1}|{\psi_s}\rangle=2\cos(k)\langle{0}|{\psi_s}\rangle$
 which represents the discontinuity introduced by the obstacle. 
 Therefore the  reflection $R=\frac{|B|^2}{|A|^2}$ and transmission $T=\frac{|C|^2}{|A|^2}$ coefficients can be easily calculated through:
\begin{equation}
           \begin{cases}
            A+B=C\\
            Ae^{-ik}+Be^{ik}=C(2\cos(k)+\Delta-e^{ik})
        \end{cases}
\longrightarrow\quad
          \begin{cases}
         B=\frac{1}{\frac{2i\sin{k}}{\Delta}-1}A\\
        C=\frac{1}{1-\frac{\Delta}{2i\sin{k}}}A
       \end{cases},
      \label{bcsca}
\end{equation}
and they have the expressions:
\begin{equation}
   R(\Delta,k)=\frac{1}{1+\dfrac{4\sin^2(k)}{\Delta^2}}, \quad T(\Delta,k)=\frac{1}{1+\dfrac{\Delta^2}{4\sin^2(k)}}.
         \label{rtsca}
\end{equation}
These coefficients closely resemble those corresponding to a delta potential in a continuous system \cite{Griffiths}; 
in particular, the coefficients only depend on $\Delta^2$, meaning that there is no difference between an 
attractive or repulsive potential for what concerns scattering.
If $\Delta$ is fixed, $T$ is maximum for $k=\frac{\pi}{2}$, which corresponds to the highest group velocity (but not to the highest energy). Consistently, at the same value of $k$, $R$ has a minimum. As the absolute value of $\Delta$ is increased, the transmission coefficient drops to smaller values, as reported in Fig. \ref{fig:trcoefficient}.
%
For every incident  $\ket{k}$ we may thus define:
\begin{align}
S\ket{k}=\frac{B}{A}\ket{-k}+\frac{C}{A}\ket{k}
\end{align}
where we introduced a scattering matrix $S$ whose elements give information on the reflection and transmission coefficients \cite{Griffiths} .
If we set $A=|A|$ and we highlight the phases of the reflected and transmitted waves, we obtain:
\begin{align}
    S\ket{k}&=
    \frac{\abs{B}}{\abs{A}}e^{i\phi_B}\ket{-k}+\frac{\abs{C}}{\abs{A}}e^{i\phi_C}\ket{k}=e^{i\phi_B}\qty(\sqrt{R(\Delta,k)}\ket{-k}+\sqrt{T(\Delta,k)}e^{i(\phi_C-\phi_B)}\ket{k}).
\end{align}
The relative phase $e^{i(\phi_C-\phi_B)}$ can be computed from the ratio $\frac{C}{B}$ from Eq. \eqref{bcsca} and is equal to $\pi/2$. It follows that: 
\begin{equation}
    S\ket{k}=e^{i\phi_B(\Delta,k)}\qty(\sqrt{R(\Delta,k)}\ket{-k}+i\sqrt{T(\Delta,k)}\ket{k}),
    \label{eqn:scattering_post}
\end{equation}
with the phase $ \phi_B(\Delta,k)=\arctan\qty(\frac{2\sin(k)}{\Delta})$.\\
\\
It is possible to define the reflection and transmission coefficients  for more general states. 
Given an initial localized wave packed $\ket{\psi_0}$ placed on the left of the obstacle, its time-evolved state is
\begin{align}
\ket{\psi(t)}=e^{-i H t}\ket{\psi_0}.\label{dynam}
\end{align}
We define the time-dependent probabilities
\begin{equation}
   \rho(t)={\sum_{j<0}\abs{\braket{j}{\psi(t)}}^2}, \qquad \tau(t)={\sum_{j>0}\abs{\braket{j}{\psi(t)}}^2}, \qquad \delta(t)=\abs{\braket{0}{\psi(t)}}^2.
    \label{eqn:scattering_coefficients_def}
\end{equation}
The  quantities $\rho(t)$ and $\tau(t)$ are indeed the probability of finding the  walker before and after the obstacle, respectively.
The defect coefficient $\delta(t)$ is the remaining probability, namely the probability of finding the particle on the obstacle site.
In particular, when the scattering is over, 
the coefficient $\delta(t)$ is expected to vanish  and consequently  $\rho(t)+\tau(t)=1$.
%
%
%
\subsection{Gaussian wavepackets}
The vector described by Eq. \eqref{eqn:scattering_post} is the mathematical building block from which we derive the asymptotic values of quantities of interest, however it is not normalizable and does not represent a physical state. For this reason, we now  introduce more realistic states that are spatially localized. In particular, we consider a discretized version of a Gaussian wave packet:
\begin{equation}
    \ket{\Gaus_{\kk}}=\NC \sum_{j\in \mathbb{Z}}e^{-\frac{(j-\mu)^2}{2\sigma^2}}e^{i\kk j}\ket{j}.
    \label{eqn:gausspos}
\end{equation}
The probability distribution of this state is a discretized Gaussian function with mean $\mu$ and variance $\frac{\sigma^2}{2}$. $\mathcal{N}$ is a normalization constant while the parameter $\kk\in(-\pi,\pi]$ represents the mean of the probability distribution in the momentum basis.
The $\ket{\Gaus_{\kk}}$  state in  momentum basis is still  Gaussian under proper assumptions and it  has the expression:
\begin{align}
  & \ket{\Gaus_{\kk}}=   \int_{-\pi}^{\pi} g_{\kk}(k)\ket{k}\,dk,
  \label{gmom}\\
   \text{with}\quad
    & g_{\kk}(k)=\braket{k}{\Gaus_{\kk}}\approx \sqrt{\frac{\sigma}{\pi^{1/2}}} \, e^{-\frac{(k-\kk)^2\,\sigma^2}{2}}e^{-i\mu k}.
    \label{eqn:gaussmom}
\end{align}
The detailed derivation of expression \eqref{eqn:gaussmom} is shown in Appendix \ref{appgaus}.\\
The crucial approximation  made to obtain this expression is to consider narrow wave packets in the reciprocal space.
Therefore, the Fourier transform of the Gaussian wave packet is not exactly a Gaussian  in the momentum basis.
Nevertheless, if the transformed state is sufficiently localized in reciprocal space, Eq.   \eqref{eqn:gaussmom} is a reasonable approximation. 
%
%
\subsection{Scattering with Gaussian wave packets}
Here, we want to analyze the asymptotic scattering 
properties of an incident Gaussian wave packet. In order to do so, we exploit the results obtained for single momentum states $\ket{k}$.
The Gaussian state  in the momentum basis has the expression \eqref{gmom}
where the Gaussian weights have been included in $g_{\kk}(k)$.\\
We consider a wave packet incident on the obstacle from the left ($j<0$).
Using \eqref{eqn:scattering_post} and linearity, the scattered Gaussian state can be written in the asymptotic limit as:
\begin{align}
    \ket{\psi_{\kk,\Delta}}&=S\ket{\Gaus_{\kk}}=\int_{-\pi}^{\pi}g_{\kk}(k)  S\ket{k}\,dk\nonumber\\
    &=\int_{-\pi}^{\pi}g_{\kk}(k)e^{i\phi_B(\Delta,k)}\qty(\sqrt{R(\Delta,k)}\ket{-k}+i\sqrt{T(\Delta,k)}\ket{k})\,dk\nonumber\\
   &= \int_{-\pi}^{\pi}\qty(e^{-i\phi_B(\Delta,k)}\sqrt{R(\Delta,k)}|g_{-\kk}(k)|e^{i\mu k}+e^{i\phi_B(\Delta,k)}i\sqrt{T(\Delta,k)}|g_{\kk}(k)|e^{-i\mu k})\ket{k}\, dk,
    \label{eqn:gaussianpost2}
    \end{align}
%
where, in the last line, we used the equalities  $|g_{\kk}(-k)|=|g_{-\kk}(k)|$, $R(\Delta, k)=R(\Delta,-k)$ and $\phi_B(\Delta,k)=-\phi_B(\Delta,-k)$.
By inspection of Eq. \eqref{eqn:gaussianpost2} we learn that  the original Gaussian wave packet is divided into the superposition of two wave packets centered around opposite values of momentum $k_0$ and $-k_0$, corresponding to the   transmitted and  reflected wave function respectively. These two wave packets are not Gaussian anymore, since they are weighted with scattering coefficients that depend on $k$.
It is important to highlight that this description fails if the two wave packets overlap, which  
 can happen  if the original state is spread in $k$-space or if its mean is $\kk \approx 0$ (or any multiple of $\pi$).
Assumption of a narrow initial wave packet in $k$-space was already imposed in order to derive Eq. \eqref{eqn:gaussmom} 
while asking for a  $\kk \neq 0$  corresponds to considering a wave packet with group velocity different from  zero.
\begin{figure}[t]
    \centering
        {\includegraphics[width=0.95\textwidth]{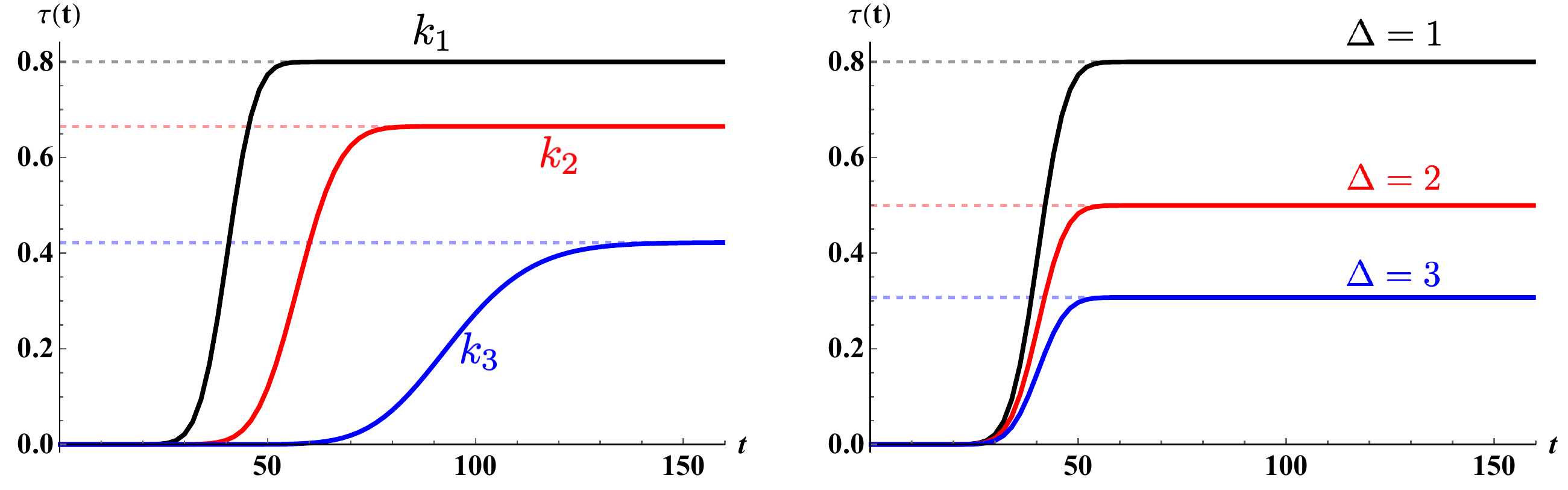}}
    \caption{Transmission probability $\tau(t)$. The left plot is for a fixed value of $\Delta=1$ and for decreasing values of $k_0=k_1,\, k_2, \,k_3$, with $k_1=1.6$  (black), $k_2=0.78$  (red), $k_3=0.44$  (blue). In the right plot, $k_0=1.6$ is kept fixed while varying the disorder $\Delta=1$ (black), $\Delta=2$ (red), $\Delta=3$ (blue). The dashed lines corresponds to the value of the transmission coefficient $\tau_{\mathcal{G}}(k_0,\Delta)$ in Eq. \eqref{eqn:scattering_coefficients_gaussian}. In both plots we considered $\sigma=15$.
    } 
      \label{fig:gaust}
\end{figure}
With these assumptions, the transmission and reflection coefficients can be calculated 
considering  the probabilities of the reflected and transmitted wave packets
\begin{equation}
    \rho_{\Gaus}(k_0, \Delta)=\int_{-\pi}^{\pi}R(\Delta,k)\abs{g_{\kk}(k)}^2\,dk, \quad \tau_{\Gaus}(k_0, \Delta)=\int_{-\pi}^{\pi}T(\Delta,k)\abs{g_{\kk}(k)}^2\,dk.
    \label{eqn:scattering_coefficients_gaussian}
\end{equation}
This results are confirmed by numerical evaluation of the $\rho(t)$ and $\tau(t)$ coefficients in Eq. \eqref{eqn:scattering_coefficients_def} and shown in Figure \ref{fig:gaust}. The dynamics of the walker is computed thought Eq. \eqref{dynam} for fixed values of $k_0$ and $\Delta$.
The figure shows that at long times, i.e. in the asymptotic limit, the transmission probability achieves exactly $\tau_{\Gaus}(k_0, \Delta)$. 
A large transmission probability is associated with  high values of $k_0$ and small values of $\Delta$, 
while a small initial central momentum and a large barrier prevent  good transmission. 
\section{Quantum estimation of a scattering potential}\label{est}
After having derived the scattered expression of a Gaussian wave packet, 
we  turn our attention to the optimal estimation of the barrier height, i.e. of the parameter $\Delta$. In order to do so, we prepare an initial Gaussian wave packet with initial central momentum $k_0$.
In a scattering experiment, measurements can be performed only on the scattered state, which has the expression of Eq.  \eqref{eqn:gaussianpost2}, that we report here for convenience:
\begin{equation}
    \ket{\psi_{\kk,\Delta}}=
    \int_{-\pi}^{\pi}g_{\kk}(k)e^{i\phi_B(\Delta,k)}\qty(\sqrt{R(\Delta,k)}\ket{-k}+i\sqrt{T(\Delta,k)}\ket{k})\,dk.\nonumber
\end{equation}
In order to compute the QFI, Eq. \eqref{eqn:qfipure}, we need the derivative:
\begin{align}
    & \ket{\partial_\Delta \psi_{\kk,\Delta}}=\int_{-\pi}^{\pi}g_{\kk}(k) e^{i\phi_B(\Delta,k)}
  \times\nonumber\\
  &\quad\times \left[i\partial_\Delta\phi_B(\Delta,k) \qty(\sqrt{R(\Delta,k)}\ket{-k}+i\sqrt{T(\Delta,k)}\ket{k})
     +\qty(\frac{\partial_\Delta R(\Delta,k)}{2\sqrt{R(\Delta,k)}}\ket{-k}+i\frac{\partial_\Delta T(\Delta,k)}{2\,\sqrt{T(\Delta,k)}}\ket{k})\right]dk,\nonumber
\end{align}
and the inner products:
\begin{align}
        \braket{\partial_\Delta \psi_{\kk,\Delta}}{\partial_\Delta \psi_{\kk,\Delta}}&=
      \int_{-\pi}^{\pi}  \abs{g_{\kk}(k)}^2\left( [\partial_\Delta \phi_B(\Delta,k)]^2 +
      \frac{[\partial_\Delta R(\Delta,k)]^2}{4\,R(\Delta,k)}+\frac{[\partial_\Delta T(\Delta,k)]^2}{4\,T(\Delta,k)}\right) dk\\
 \braket{\psi_{\kk,\Delta}}{\partial_\Delta \psi_{\kk,\Delta}}&=i \int_{-\pi}^{\pi}
 \abs{g_{\kk}(k)}^2 \,\partial_\Delta\phi_B(\Delta,k) \,dk,
\end{align}
with  $\partial_\Delta R(\Delta,k)+ \partial_\Delta T(\Delta,k)=0$.
We remind the reader  that in this work we are always assuming that
 the reflected and transmitted wavepackets of the post-scattering state do not overlap, neither in position nor in momentum space.
Notice that with this assumption we also exclude slow states, i.e. those states with $\kk \approx 0 $ or $\kk \approx \pi$. 
The QFI for an initial  Gaussian wave packet may be 
computed through Eq. \eqref{eqn:qfipure}:
\begin{align}
    H_{\Gaus}(k_0, \Delta)&=\int_{-\pi}^{\pi} \abs{g_{\kk}(k)}^2 \qty( \frac{[\partial_\Delta R(\Delta,k)]^2}{R(\Delta,k)}+\frac{[\partial_\Delta T(\Delta,k)]^2}{T(\Delta,k)}+4  [\partial_\Delta \phi_B(\Delta,k)]^2  )dk\nonumber\\
   &\quad -4\qty(\int_{-\pi}^{\pi}\abs{g_{\kk}(k)}^2 \partial_\Delta\phi_B(\Delta,k)dk)^2    \label{qfi}\\ 
   &= \frac{16 \sin^2 k_0}{[2+\Delta^2 - 2 \cos (2 k_0)]^2} + \frac{g_H(k_0,\Delta)}{\sigma^2} + O(1/\sigma^3)\,,
   \label{qfi2}
\end{align}
where the explicit expression of $g_H(k_0,\Delta)$ is reported in Appendix \ref{thegs}. 
\begin{figure}[t]
    \centering
        {\includegraphics[width=0.99\textwidth]{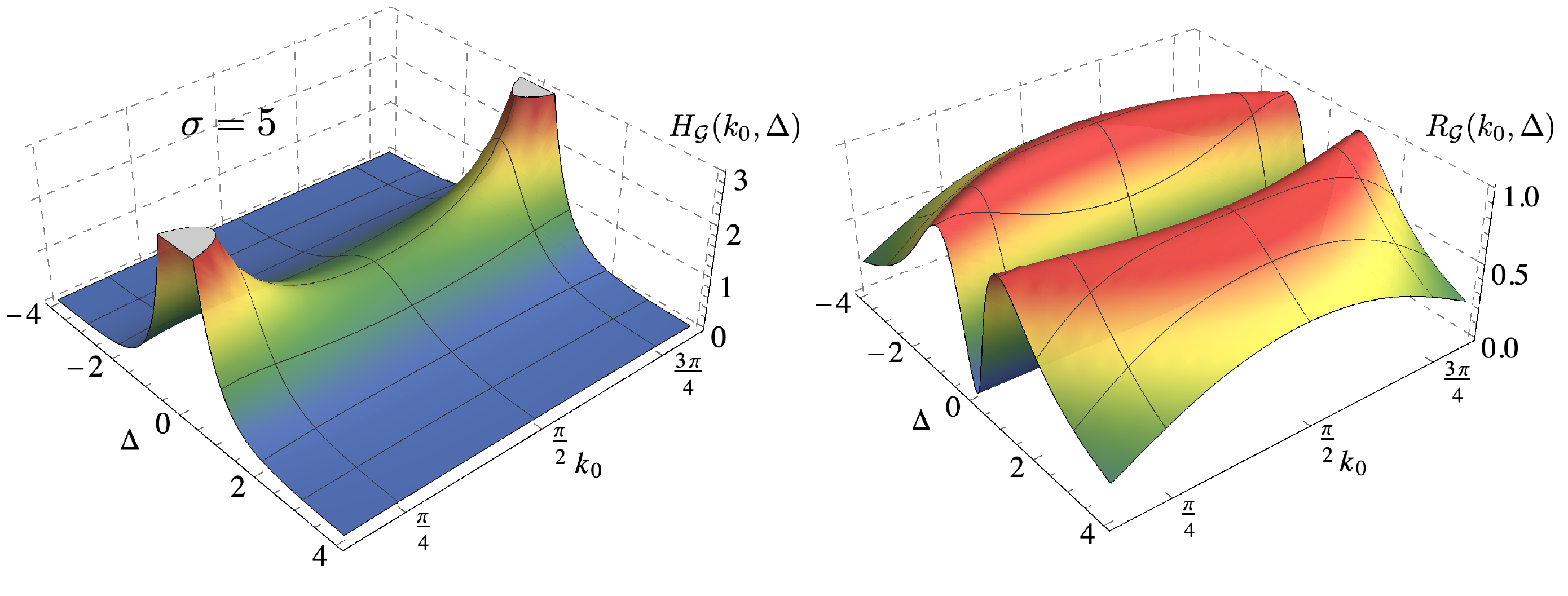}}
    \caption{ Left: QFI $H_{\mathcal{G}}(k_0,\Delta)$ for an initial Gaussian wave packet with $\sigma=5$. Right: QSNR $R_{\mathcal{G}}(k_0,\Delta)$ for the same initial Gaussian wavepacket. } 
      \label{fig:3}
\end{figure}
\begin{figure}[t]
    \centering
        {\includegraphics[width=0.99\textwidth]{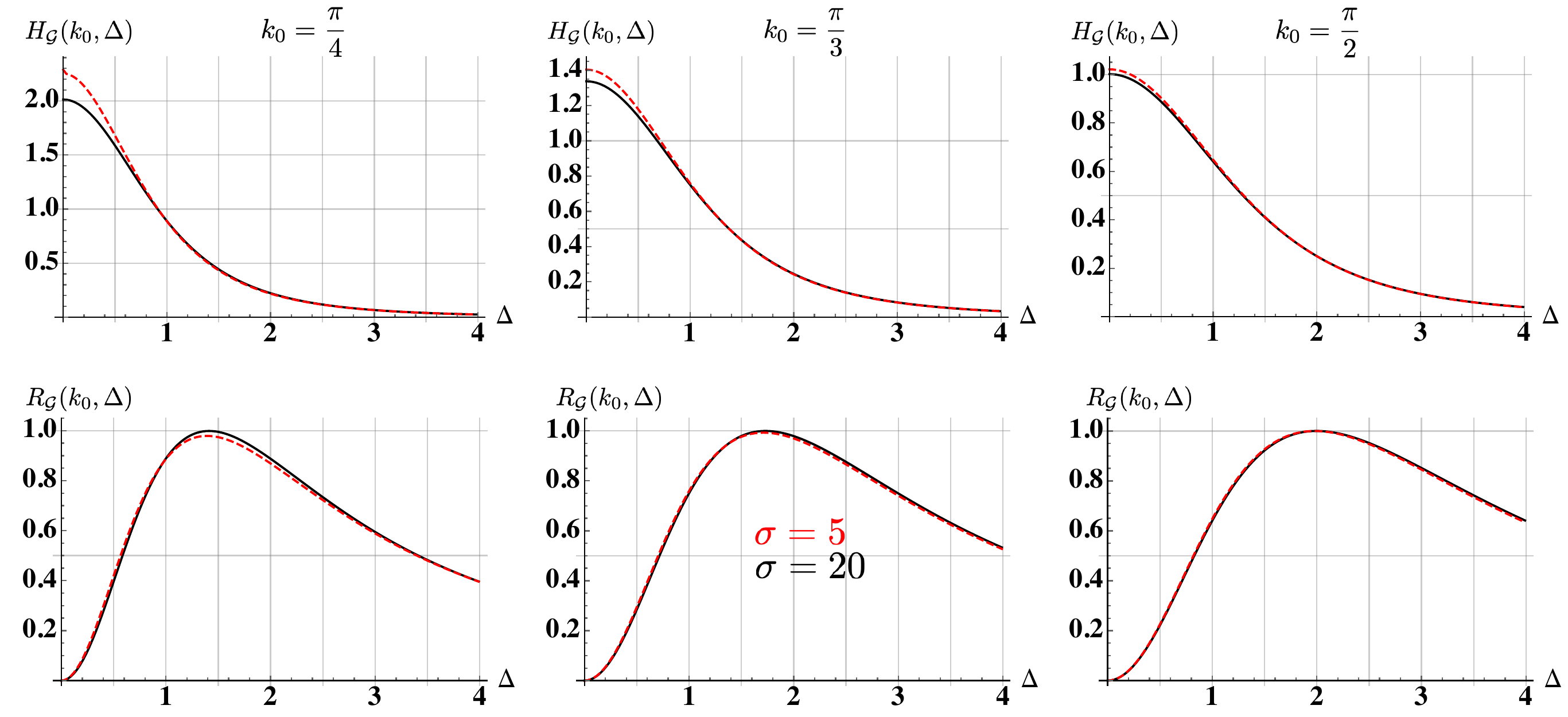}}
    \caption{  Comparison between the QFI (upper panel) and the QSNR (lower panel)  with a large and a narrow initial wave-packet in $k$-space, as a function of $\Delta$ and for three different values of $k_0$. 
 The black solid lines are for $\sigma=20$ while the dashed red lines are for $\sigma=5$. The considered values of  initial momentum are   $k_0=\frac{\pi}{4},\frac{\pi}{3},\frac{\pi}{2}$ for the left, center and right column respectively. } 
      \label{fig:4}
\end{figure}
%
The typical behavior of the QFI as a function of  $\Delta$ and the 
initial central momentum $k_0$ is shown in Figure \ref{fig:3}. 
Since we want to avoid overlaps of the reflected and transmitted wave functions 
 in momentum space, we exclude values for $k_0$ in the neighborhood of $k_0=0$ and $k_0=\pi$.
 The QFI is symmetric under the exchange of the sign of the barrier, i.e. $\Delta\rightarrow -\Delta$ and it has a maximum centered in $\Delta=0$. 
Small values of the barrier height $|\Delta| \ll 1$ have  a larger QFI with respect to higher barriers. The spread of the wave packet $\sigma$ affects the maximum precision only for  $|\Delta| \ll 1$,  as shown in the upper panel of Figure \ref{fig:4}. From these plots, we can also see that the initial central momentum has an important role: in fact, as $k_0$ is increased from small values to $\frac{\pi}{2}$ the maximum of the QFI decreases.
\par
In order to compare the error of an estimator with the true value of the parameter to be estimated, we also addressed the QSNR, defined in  Eq. \eqref{qsnr}. Its behavior is 
shown in the right plot of Figure \ref{fig:3} and in the lower panel of Figure \ref{fig:4}, for three different values of the initial central
momentum $k_0$. The QSNR has a maximum for $\Delta\neq 0$, which corresponds to the
value of the barrier height that can be better estimated. 
As the value of the initial central momentum is increased toward $k_0=\frac{\pi}{2}$, the value
of the optimal $\Delta$ slightly increases. The dependency on $\sigma$ is negligible when considering the QSNR, as shown in the lower plots, where the behaviors for $\sigma=5$ and $\sigma=20$ are compared. Quite remarkably, the maximum value of the QSNR is very similar, $R_{\mathcal{G}}\approx 1$ for the considered values of $k_0$, thus making the initial central momentum a tool to fine tune the optimal value of $\Delta$, but not the corresponding precision.%
\par
The behaviour of the QSNR has an intuitive and straightforward physical interpretation. If the height of the barrier is negligible ($\Delta \ll 1$) then the walker is mostly transmitted anyway and small variations of $\Delta$ itself are very difficult to detect. Similarly, if $\Delta \gg 1$ the walker is mostly reflected independently on the exact value of $\Delta$. On the other hand, for intermediate values of $\Delta$ the wavefunction of the walker is sensitive to its value, and measuring the walker indeed provides information. This picture is confirmed if one looks at the zero-th order 
expression of the QFI in Eq. (\ref{qfi2}), which says that the maxima of the QSNR are located at $\Delta^2 = 2 [1- \cos (2 k_0)]$. Notice that the values of $(\Delta,k_0)$ satisfying this relations are those making 
the reflection and transmission equal to each other $R(\sqrt{2*[1- \cos (2 k_0)]},k_0)=T(\sqrt{2*[1- \cos (2 k_0)]},k_0)=\frac12$. 
\subsection{Dichotomic position measurement}
We now address the question of whether a realistic position measurement 
is optimal, i.e. its FI equals the QFI defined in Eq. \eqref{qfi}. In particular, we consider a dichotomic measurement that just tells us if the particle is located on 
the left  or on the right side of the barrier. 
Since we know from  Eq.s \eqref{eqn:scattering_coefficients_gaussian} that the quantities $\rho_{\mathcal{G}}(k_0,\Delta)$ and $ \tau_{\Gaus}(k_0,\Delta)$ correspond 
to the probabilities of finding the particle before or after the obstacle, the FI takes the expression:
\begin{align}
    F_{\Gaus}(k_0,\Delta)&=\frac{[\partial_\Delta \rho_{\Gaus}(k_0,\Delta)]^2}{\rho_{\Gaus}(k_0,\Delta)}+\frac{[\partial_\Delta \tau_{\Gaus}(k_0,\Delta)]^2}{\tau_{\Gaus}(k_0,\Delta)} = \frac{[\partial_\Delta \tau_{\Gaus}(k_0,\Delta)]^2}{\tau_{\Gaus}(k_0,\Delta)[1-\tau_{\Gaus}(k_0,\Delta)]} 
    \label{eqn:fishgauss}
    \\
    &= \frac{16 \sin^2 k_0}{[2+\Delta^2 - 2 \cos (2 k_0)]^2} + \frac{g_F(k_0,\Delta)}{\sigma^2} + O(1/\sigma^3)\,,
    \label{eqn:fishgauss}
\end{align}
where the explicit expression of $g_F(k_0,\Delta)$ is reported in Appendix \ref{thegs}.
As the value of $\sigma$ is increased, i.e. the wave packet is more localized in 
$k$-space, the FI of the dichotomic measurement approaches the QFI. The second order coefficients $g_s(k_0,\Delta)$, $s=H,F$ are different for the QFI and the FI (see Appendix \ref{thegs}), but
in the range of parameters we have explored ($\sigma>5$, $0<\Delta\leq 4$, $0<k_0<\pi$)
the ratio $\gamma(k_0,\Delta)=F_{\Gaus}(k_0,\Delta)/ H_{\Gaus}(k_0,\Delta)$ is always larger than $\gamma(k_0,\Delta)>0.95$. We conclude that a dichotomic position measurement is nearly optimal to estimate the height of the potential barrier $\Delta$. 
\section{Conclusions}\label{concl}
In this work we have introduced and discussed a general probing scheme for scattering problems based on continuous-time quantum walks.  In particular, we have considered a one-dimensional lattice, with an impurity at its center, i.e. a potential barrier of height $\Delta$, and discussed in details how to quantify the maximum extractable information about the parameter $\Delta$. 
\par 
Using the continuous-space case as a guide for attacking the problem, we have 
first introduced the single-momentum scattered states $S\ket{k}$, and used them to
compute the reflection and transmission coefficients of the considered potential.  From the scattered states,  we   built up the asymptotic Gaussian states, i.e. physical states 
that depend, in addition to $\Delta$,  upon the initial central momentum $k_0$ and the spread of the wave packet in position space $\sigma$. We then derived the reflection and transmission probability of such wave packets. Finally, we computed the QFI for the parameter $\Delta$. 
We showed that the QFI has a maximum for $\Delta=0$ and it is strongly affected by the value of  $k_0$. In particular values of $k_0$ near $\frac{\pi}{2}$ lead to a smaller QFI. Moreover, for  $|\Delta|\ll1$, a small $\sigma$ can increase the precision of the estimation. However, inspection of the QSNR did not show a noticeable difference in its behavior depending on the value of $\sigma$ or $k_0$. The QSNR has a maximum for $\Delta \neq 0$, indicating that given the value of the central momentum $k_0$, there exists a value for $\Delta$ that can be better estimated, leading to unit QSNR 
 independently from $\sigma$ and $k_0$.
\par
Finally, we have investigated the performances of a dichotomic position measurement, that is a binary measurement that is just able to distinguish if a particle is located on the left (reflected) or on the right (transmitted) of the potential barrier. We have shown that this measurement is optimal, i.e. its FI is equal to the  QFI, for large initial wave packets (in position space), while it is nearly optimal for narrow 
initial wave packets. 
\par
Our work paves the way to the characterization of more involved forms of potentials using a single-particle continuous-time quantum walk as a probe. Extensions of this work may also include  more complex structures, such as multi-dimensional graphs, where  imperfections created during  the fabrication process need to be estimated in order to better control the quantum dynamics over such networks.

\authorcontributions{All the authors have contributed equally.}

\funding{This research received no external funding}

\acknowledgments{MGAP is member of INdAM-GNFM}

\conflictsofinterest{The authors declare no conflict of interest} 

\abbreviations{The following abbreviations are used in this manuscript:\\
\noindent 
\begin{tabular}{@{}rl}
CTQW & Continuous-time quantum walk\\
CR& Cramér-Rao\\
FI &  Fisher information\\
QFI & Quantum Fisher information\\
QSNR & Quantum signal-to-noise ratio
\end{tabular}}

\appendixtitles{yes} 
\appendix
\section{Gaussian wavepacket in $k$-space}\label{appgaus}
Consider the Gaussian wavepacket in position space defined by Eq. \eqref{eqn:gausspos}. Here we show that its expression in $k$-space, within certain approximations, is given by expression \eqref{eqn:gaussmom}.
We start by considering the the projection of Eq.  Eq. \eqref{eqn:gausspos} into a state $\ket{k}$:
\begin{equation}
    \braket{k}{\Gaus_{\kk}}=\frac{\NC}{\sqrt{2\pi}}\sum_{j\in \mathbb{Z}}e^{-\frac{(j-\mu)^2}{2\sigma^2}}e^{i(\kk-k) j}.
\end{equation}
The  infinite sum can be calculated using Poisson summation formula which states that, for suitable functions $f$:
 $   \sum_{j\in \mathbb{Z}}f(j)=\sum_{n \in \mathbb{Z}}\hat{f}(n)=\sum_{n\in \mathbb{Z}}\int_{-\infty}^{+\infty}f(x)e^{-i2\pi n x} \dd{x}$.
In our particular case:
\begin{equation}
   \braket{k}{\Gaus_{\kk}}=  \frac{\NC}{\sqrt{2\pi}}\sum_{j\in \mathbb{Z}}e^{-\frac{(j-\mu)^2}{2\sigma^2}}e^{i(\kk-k) j}=\frac{\NC}{\sqrt{2\pi}}\sum_{n \in \mathbb{Z}}\int_{-\infty}^{+\infty}e^{-\frac{(x-\mu)^2}{2\sigma^2}}e^{i(\kk-k) x}e^{-i2\pi n x} \dd{x}.\label{appuno}
\end{equation}
The last integral is a continuous Fourier transform of a Gaussian function, therefore:
\begin{equation}
    \int_{-\infty}^{+\infty}e^{-\frac{(x-\mu)^2}{2\sigma^2}}e^{i(\kk-k) x}e^{-i2\pi n x} \dd{x}=\sqrt{2\pi \sigma^2}e^{-\frac{(2\pi n+k-\kk)^2}{2\frac{1}{\sigma^2}}}e^{-i\mu(2\pi n+k-\kk)}.\label{appint}
\end{equation}
Inserting Eq. \eqref{appint} into \eqref{appuno} (discarding the constant global phase $e^{i\mu \kk}$), we obtain:
\begin{equation}
    \braket{k}{\Gaus_{\kk}}=\NC \sigma \sum_{n\in \mathbb{Z}}e^{-\frac{(2\pi n+k-\kk)^2}{2\frac{1}{\sigma^2}}}e^{-i\mu(2\pi n+k)}.
\end{equation}
The transformed state is not a Gaussian state but it is an infinite sum of Gaussian states periodically displaced. 
However, if the wavepacket is localized enough in reciprocal space, it is possible to approximate the last infinite summation by keeping only the central term $n=0$ (it is always possible to shift the definition of $k$ and $\kk$ in the interval $[-\pi,\pi)$ because they are defined modulo $2\pi$).
The localization assumption is needed in order to consider only one term, otherwise  the tails of adjacent Gaussian functions could overlap. With this assumption:
\begin{equation}
    \Gaus_{\kk}(k)=\braket{k}{\Gaus_{\kk}}\approx\NC \sigma e^{-\frac{(k-\kk)^2}{2\frac{1}{\sigma^2}}}e^{-i\mu k}.
\end{equation}
Thus, a discrete Gaussian state in the position basis remains a Gaussian state in reciprocal space within the considered approximation.
The calculation of the normalization constant $\NC$ reduces to the calculation of a Gaussian integral:
\begin{align}
    1&=\int_{-\pi}^{\pi}\abs{\Gaus_{\kk}(k)}^2 \dd{k}\approx \int_{-\infty}^{\infty}\abs{\Gaus_{\kk}(k)}^2 \dd{k}=\abs{\NC}^2\sigma \sqrt{\pi},
\end{align}
with
\begin{align}
       \abs{\NC}^2&\approx\frac{1}{\sqrt{\pi \sigma^2}}.
\end{align}

\section{The explicit expression of the functions $g_H(\Delta,k_0)$ and $g_F(\Delta,k_0)$}
\label{thegs}
We have
\begin{align}
g_H (k_0,\Delta) & =  \frac{4\,\big[3 \cos 6 k_0 + 2 (5 \Delta^2 -1) \cos 4 k_0 + 3 (3 \Delta^4-19) \cos 2k_0 + \Delta^4-10 \Delta^2+18\big]}{\big[\Delta^2+2(1-\cos 2k_0)\big]^4} \,,  \\
g_F (k_0,\Delta) & =   \frac{8\,\big[\cos 6 k_0 + 6\Delta^2  \cos 4 k_0 +  (\Delta^4-9) \cos 2k_0 -6 \Delta^2+8\big]}{\big[\Delta^2+2(1-\cos 2k_0)\big]^4} \,.
\end{align}
\reftitle{References}

\externalbibliography{yes}
\bibliography{scattering}

\end{document}